\documentclass[conference]{IEEEtran}
\IEEEoverridecommandlockouts
\usepackage{cite}
\usepackage{amsmath,amssymb,amsfonts}
\usepackage{algorithmic}
\usepackage{graphicx}
\usepackage{textcomp}
\usepackage{xcolor}
\usepackage{hyperref}
\usepackage{listings}
\usepackage{xcolor}
\usepackage{array}
\def\BibTeX{{\rm B\kern-.05em{\sc i\kern-.025em b}\kern-.08em
    T\kern-.1667em\lower.7ex\hbox{E}\kern-.125emX}}
\begin{document}

\title{\textsc{MOSAIK} 3.0: Combining Time-Stepped and Discrete Event Simulation\\

\thanks{Funded by the Lower Saxony Ministry of Science and Culture under grant number 11-76251-13-3/19 - ZN3488 within the Lower Saxony “Vorab“ of the Volkswagen Foundation and supported by the Center for Digital Innovations (ZDIN).
\\
\copyright2024 IEEE.~https://doi.org/10.1109/OSMSES54027.2022.9769116
Personal use of this material is permitted. Permission from IEEE must be obtained for all other uses, in any current or future media, including reprinting/republishing this material for advertising or promotional purposes, creating new collective works, for resale or redistribution to servers or lists, or reuse of any copyrighted component of this work in other works.}

}

\author{\IEEEauthorblockN{Annika Ofenloch, Jan Sören Schwarz, Deborah Tolk, Tobias Brandt, \\ Reef Eilers, Rebeca Ramirez, Thomas Raub, Sebastian Lehnhoff}
\IEEEauthorblockA{\textit{R\&D Division Energy} \\
\textit{OFFIS Institute}\\
Oldenburg, Germany \\
annika.ofenloch@offis.de}
}

\maketitle

\begin{abstract}
Co-simulation is commonly used for the analysis of complex cyber-physical energy systems (CPES). Different domain-specific simulation tools and modeling approaches are used to simulate all or parts of the system. The co-simulation framework \textsc{mosaik} is a powerful tool to couple these simulation tools and models. This paper identifies the limitations of \textsc{mosaik}~2 for simulating systems that combine continuous and discrete behavior, and introduces the new version \textsc{mosaik}~3.0 with improved event capabilities to efficiently combine time-stepped and discrete event simulation. Here it is explained how these extensions and new features of \textsc{mosaik} can be applied and implemented for extended co-simulation scenarios.
\end{abstract}

%https://www.ieee.org/content/dam/ieee-org/ieee/web/org/pubs/taxonomy_v101.pdf

\section{Introduction} 

Energy systems are composed of a large number of subsystems, which are connected to and influenced by other systems---forming a highly complex dynamic cyber-physical energy system (CPES). To simulate CPES, these subsystems must be adequately modeled and either integrated into a joint model or coupled in a time-synchronous manner, which is the goal of co-simulation. As numerous models, simulation tools, and experts of different domains are available, co-simulation is a powerful tool to investigate various scenarios~\cite{Schloegl2015a}.

Complex co-simulation studies lead to the interaction of various simulation paradigms such as continuous simulation and discrete event simulators, using different modeling approaches. For a power system, it is common to use continuous or discrete time modeling to simulate its dynamic behavior. Discretization and interpolation methods can be used to convert continuous time models into discrete time models that can be easily simulated in a time-stepped simulation with discrete time steps. In contrast, communication systems are typically modeled as discrete event systems, using events that are unevenly distributed over time~\cite{Mets2014,Cremona2017}.

To provide the capabilities to couple simulators with different simulation paradigms, an extended version of the co-simulation framework \textsc{mosaik}\footnote{\url{https://mosaik.offis.de}} is introduced that combines time-stepped and discrete event simulation, following the approach of hybrid co-simulation by Cremona et al. \cite{Cremona2017}. In support of hybrid co-simulation, extensions to \textsc{mosaik}'s component API and scheduling algorithm are introduced, including discrete event capabilities, superdense time feature, external events, and the explicit definition of a time resolution in a co-simulation scenario.

A general overview of the state of the art of hybrid co-simulation is given in Section~\ref{sec:state_of_art}. Specifically, Section~\ref{sec:hybrid} provides examples and insights into the new features and enhancements of \textsc{mosaik} 3.0 to support hybrid co-simulation. Lastly, we provide a summarizing discussion on future work in Section~\ref{sec:future_work} and conclude in Section~\ref{sec:conclusion}.

\section{State of The Art} \label{sec:state_of_art} 
The co-simulation framework \textsc{mosaik} has been developed for years as an universal co-simulation framework for various domains~\cite{Schloegl2015a,Steinbrink2019}. However, its focus lies on usability and flexibility of co-simulation in context of the Smart Grid.
\textsc{mosaik} is based on a common interface independent of programming languages (the component API), which can be used to integrate components into \textsc{mosaik}, providing the basis for a modular and extendable simulation environment. The simulation of the scenario is defined by the scenario API and coordinated by \textsc{mosaik}'s scheduler, which synchronizes the simulators and handles the data flow between them.
\par
Although already supported by \textsc{mosaik} 2, discrete event simulations could not be implemented in a native and efficient way, as simulation components could only schedule steps for themselves. In order to not miss any potential message from other components, \textsc{mosaik} 2 simulators had to step themselves for every simulation step. Also, the provision of output was obligatory so that the absence of output (e.g., no message arriving within a communication simulation) had to be indicated via the exchange of empty data containers~\cite{Kosek2014}.
\par
This led to performance problems within certain simulation scenarios, e.g., CPES simulations considering communication systems. Here, a very high time resolution is needed to accurately model the interaction between communicating components like co-operating agents. In such scenarios, the probability of unnecessary simulator self-stepping is very high, so that there is a high potential to improve the overall co-simulation performance. Thus, \textsc{mosaik} has been extended to resolve the aforementioned inefficiencies and to allow native discrete event simulation as explained in Section~\ref{sec:hybrid}.
\par
Besides \textsc{mosaik}, many other co-simulation frameworks for Smart Grids exist.
As shown in \cite{Vogt2018}, many of them already provide discrete event functionality to some extent.
Nevertheless, in the comparison of Vogt et al. \textsc{mosaik} is also described as a discrete event-based framework, which indicates that an ambiguous definition was used for the categorization and other tools may exhibit similar implementations as \textsc{mosaik} 2.
More generic co-simulation frameworks like Ptolemy II \cite{Ptolemy1999} (e.g., used in the Building Controls Virtual Test Bed \cite{Wetter2011}) and the High-Level Architecture standard (HLA) \cite{HLA1997} (e.g., used in C2WT-TE \cite{Neema2016a}) also rely on complex synchronization methods including sophisticated discrete event simulation.
Nevertheless, due to more extensive features, these tools are more complex and less flexible to use compared to \textsc{mosaik} as stated in \cite{Steinbrink2017a}.
\par
The hybrid co-simulation approach 
references the Functional Mock-Up Interface (FMI) as an important standard for the definition of co-simulation components, which can be shared as Functional Mock-Up Units (FMU).
These FMUs do not contain a complete simulation environment, but can be integrated in \textsc{mosaik} and many other co-simulation frameworks.
FMI 2 \cite{ModelicaAssociationProjectFMI2019} does not fully support hybrid co-simulation yet \cite{Cremona2017}, but the upcoming version 3 will improve support by introduction of multiple new features, such as the establishment of \texttt{Clocks} for improved event handling, 
early return from FMU step execution in case of an event and making available super dense time not only for FMI-ME (model exchange) but also FMI-CS (co-simulation)~\cite{FMI3}.
\section{Hybrid Co-Simulation in \textsc{mosaik} 3.0}\label{sec:hybrid}
Hybrid co-simulation allows the simulation of CPES to integrate continuous time models (e.g., mechanical models, fluid dynamics models) as well as discrete event models (e.g., communication networks, software agents). Therefore, the support of hybrid co-simulations is highly relevant for the analysis of CPES.

The following extensions to the scheduling algorithm and component API support hybrid co-simulations with improved discrete event capabilities.
The improved scheduling algorithm supports both time-based and event-based simulations as well as a combination of both paradigms.
The extensions to the component API include a simulation component type, a new parameter \texttt{max\_advance} to improve the performance and prevent causality errors, and the definition of a global time resolution.
\subsection{Component type} 
\begin{figure}[!h]
\centering
\includegraphics[trim=3cm 4cm 9cm 3cm, clip, width=0.47\textwidth]{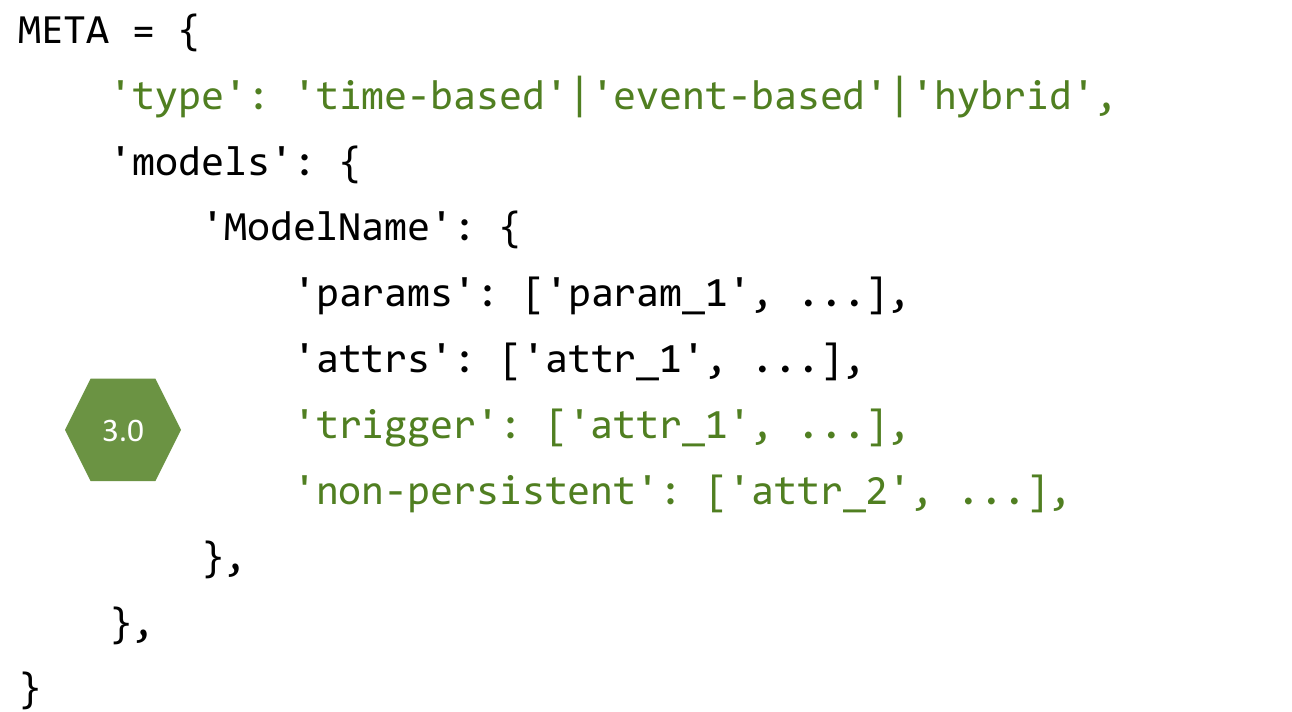}
\caption{Meta data dictionary of the simulator (new elements of \textsc{mosaik} 3.0 in green)~\cite{Raub2021}.}
\label{meta-data}
\end{figure}

As shown in Figure~\ref{meta-data}, the type of a component has been added to the meta data, which is returned by the \texttt{init} function of the API. It defines how the simulator advances through time. The \textit{time-based} simulators are the traditional ones of \textsc{mosaik} 2 with self-stepping and persistent data.
This means the output data is valid until the next simulator step, e.g., for active power of a photo-voltaic module. In some simulations, the state of the system can change at a specific time, i.e., sending/receiving of messages in a communication simulation. For those simulators, the \textit{event-based} type is introduced. The simulator is stepped as soon as there is new data available and the data is only valid for a single time step (non-persistent or transient). As the processing of an external event may need to be modeled over several time steps, \textit{event-based} simulators can optionally still trigger themselves. 

\par
The \textit{hybrid} type combines the behavior of \textit{time-based} and \textit{event-based} simulators. It allows to define persistence and triggering on attribute level, i.e., the inputs and outputs of the simulators.
Attributes, which should only be triggered by available data, have to be added to the \texttt{trigger} list. 
The output attributes are persistent as in the \textit{time-based} components, if they are not added to the \texttt{non-persistent} list.
An overview of the simulation component types is given in Table \ref{tab1}. 
{\renewcommand{\arraystretch}{1.2}% for the vertical padding
\begin{table}[htbp]
\caption{New simulation component types}
\begin{center}
\begin{tabular}{|p{1.4cm}|>{\raggedright\arraybackslash}p{1.5cm}|>{\raggedright\arraybackslash}p{1.5cm}|>{\raggedright\arraybackslash}p{2.5cm}|}
\hline
\textbf{Type} & Time-based & Event-based & Hybrid \\
\hline
\textbf{Next step} & obligatory & optional & optional \\
\hline
\textbf{Input}\hspace{0.3cm}\textbf{trigger} & never & always & only if specified \\
\hline
\textbf{Output}\hspace{0.3cm}\textbf{validity} & until next step & only for specified time & until next step if not specified as non-persistent \\ 
\hline
\end{tabular}
\label{tab1}
\end{center}
\end{table}
}

\begin{figure}[!h]
\centering
\includegraphics[trim=5cm 4.1cm 5cm 5.9cm,clip, width=0.47\textwidth]{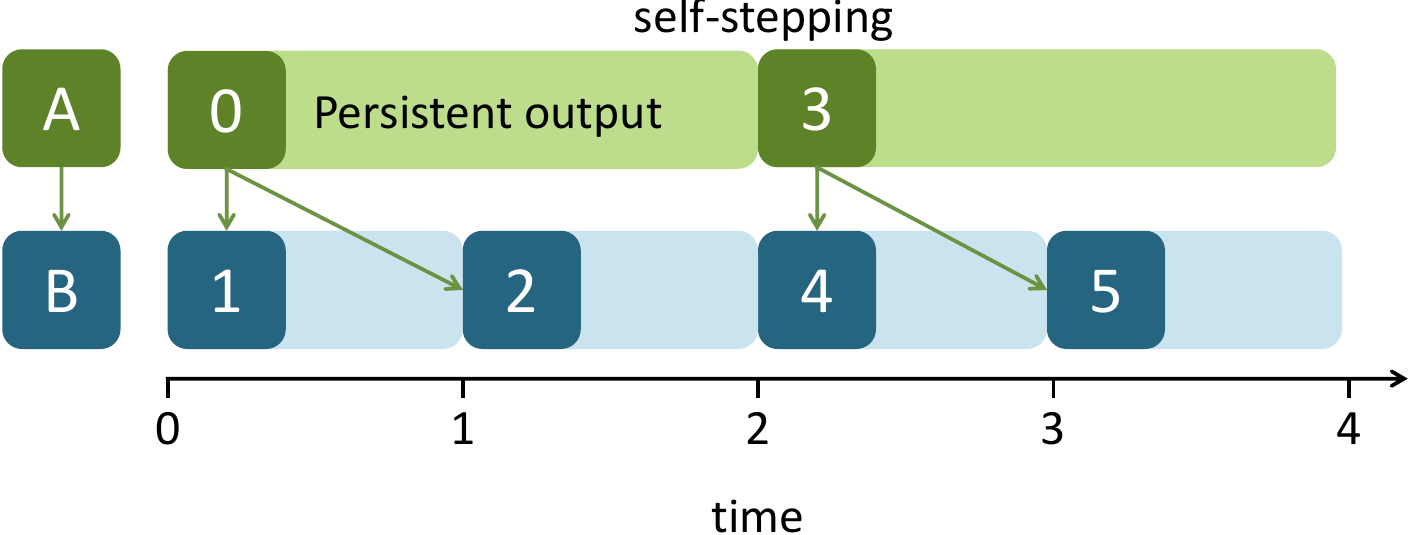}
\caption{Schematic execution of the time-based simulators A and B~\cite{Raub2021}.}
\label{time-based}
\end{figure}

As depicted in Figure~\ref{time-based}, all \textit{time-based} and \textit{hybrid} simulators start the simulation at time 0. The output of simulator A is persistent and simulator B will use the last output, even if it is at a different time step.
The x-axis shows the simulation time and the number in the colored boxes represents the order of execution.
\par
\begin{figure}[!h]
\centering
\includegraphics[trim=5cm 4.2cm 8cm 6.5cm,clip, width=0.47\textwidth]{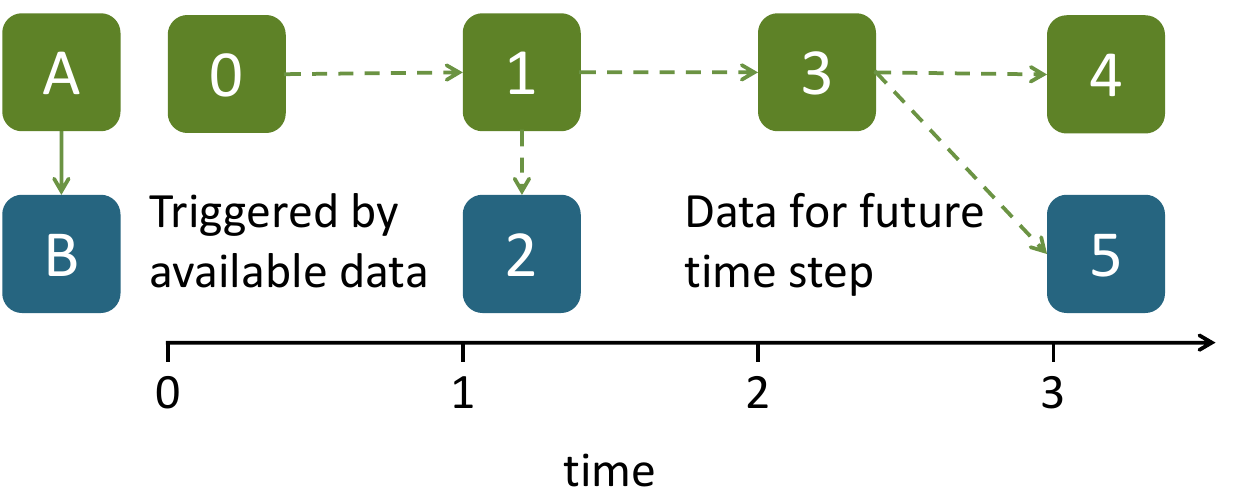}
\caption{Schematic execution of event-based simulators A and B~\cite{Raub2021}.}
\label{mosaik_des}
\end{figure}

The stepping through time is different for \textit{event-based} simulators. They do not automatically start at time 0, but whenever their first event is scheduled.
In Figure~\ref{mosaik_des}, simulator B is \textit{event-based}. Therefore, there is no need to specify when it is executed, but it will be triggered as soon as simulator A has new data for B. In other words, the simulator is stepped when an event is created, which can be also done via the \texttt{step} function's return value. By default, the data output is valid for the simulation time of the current step and triggers immediately the dependent simulators.
Alternatively, the validity of the output values can be set in the data object \texttt{outputs} via a \texttt{time} entry, e.g., for triggering depended simulators in the future.

\subsection{Max Advance} 
For performance improvements, the extended \textsc{mosaik} API features a new parameter \texttt{max\_advance}, which is included in the \texttt{step} function and determined for each time step by analyzing the simulation progress of the direct and indirect dependencies (ancestors) within \textsc{mosaik}'s environment. This parameter provides the simulator with the information how far it can advance in time without expecting new inputs from other simulators. Thus, the simulator can progress until \texttt{max\_advance} without being interrupted by \textsc{mosaik} and risking any causality errors. For instance, a communication system with a changing internal state can calculate larger time intervals as no interactions with other simulators are scheduled during that time. In addition, the simulator can stop the stepping at any intermediate fundamental time step of the interval (e.g., the message has reached its destination) to trigger new interactions with other simulators.

\begin{figure}[!h]
\centering
\includegraphics[trim=7cm 5cm 8cm 3.5cm,clip, width=0.47\textwidth]{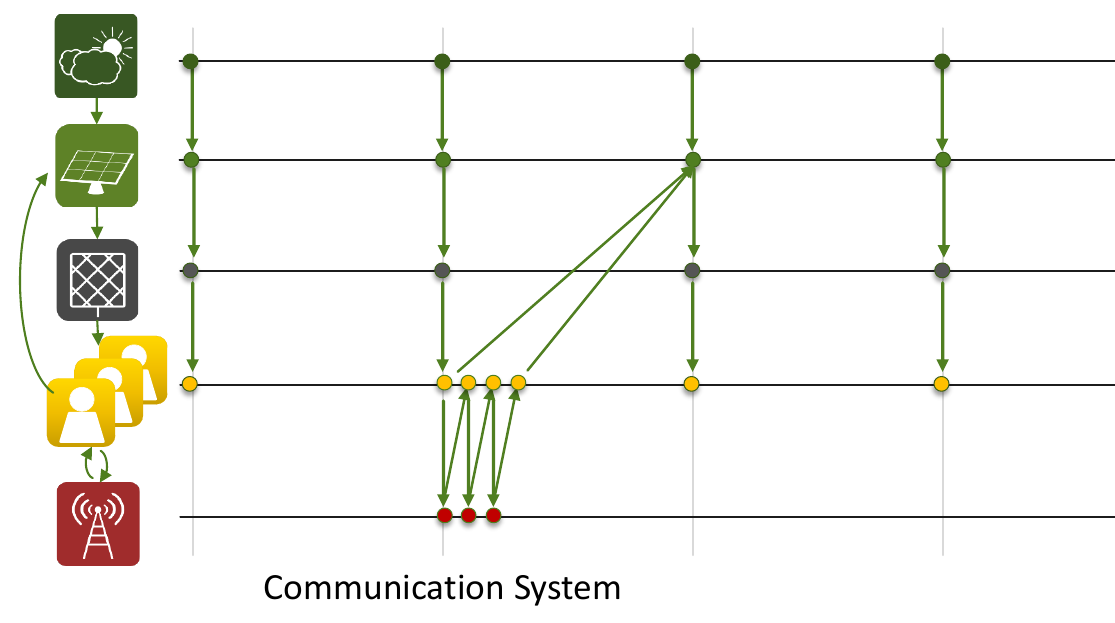}
\caption{Use Case: Integration of a communication simulation~\cite{Raub2021}.}
\label{integration-com-sim}
\end{figure}

Figure~\ref{integration-com-sim} shows the integration of a communication simulation as a use case for a discrete event simulation and the time parameter \texttt{max\_advance}.
In this setup, multiple software agents communicate with each other while the message dispatch is simulated with a modeled communication network. Here, the communication simulation can advance until \texttt{max\_advance} within a single \texttt{step} function call of \textsc{mosaik}, calculating larger time intervals without being interrupted.

Therefore, with the discrete event capabilities and the new time parameter \texttt{max\_advance}, it is more efficient and accurate to couple simulators that operate on different time scales, e.g., communication between agents in milliseconds, quasi steady state of electric grid in seconds, and photovoltaic simulations in minutes.

\subsection{Time-resolution} 
To configure a global time resolution within the scenario definition, the API was extended by a new parameter \texttt{time\_resolution}. This parameter can be set as part of the instantiation of \texttt{mosaik.scenario.World}, which stores all data and states that belong to the scenario and its simulation. Each simulator receives the time resolution as a keyword argument via the \texttt{init} function of the API, which can be handled internally by the simulator to flexibly adapt to different time resolutions. If no global time resolution is defined, the default value of \texttt{time\_resolution} is set to \texttt{1.0} (one second per \textsc{mosaik}'s internal integer time step).

For example, a time resolution of \texttt{1.0} means that \textsc{mosaik}’s internal time of \texttt{1} represents one second. If the global time resolution is set to \texttt{0.001} instead, it means that \textsc{mosaik}’s internal time of \texttt{1} represents one millisecond. In other words, this tells the simulator how to interpret \textsc{mosaik}'s \texttt{time} attribute in the \texttt{step} function call and allows the simulators to adjust their model behavior accordingly. In this way, simulators with internal time dependencies (e.g., a controller dead time) can directly adapt to various simulation scenarios.
\subsection{Superdense time} 
\par
\begin{figure}
\centering
\includegraphics[trim=8cm 4.5cm 8cm 5.5cm, clip, width=0.42\textwidth]{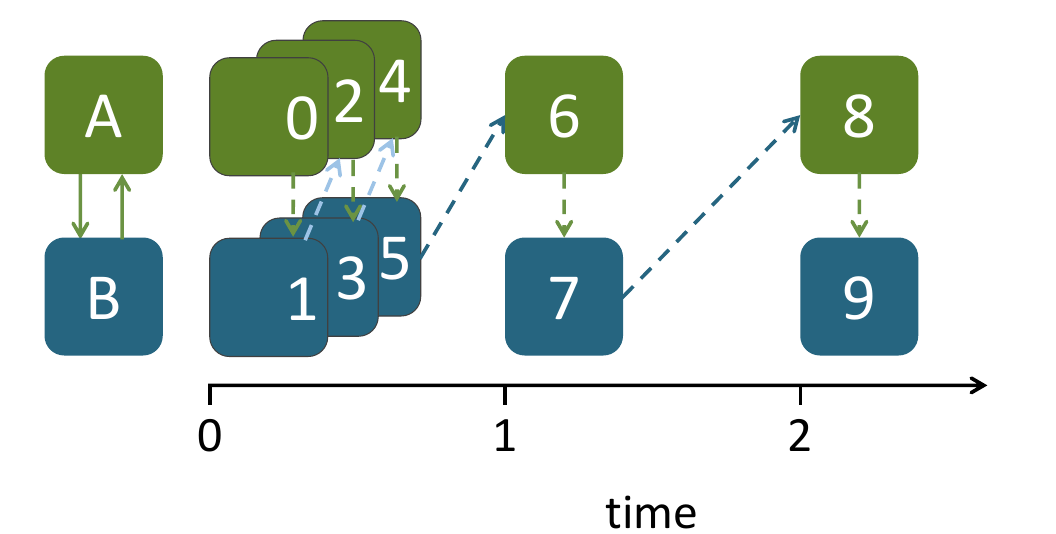}
\caption{Visualization of superdense time (i.e., same time loops) in \\\textsc{mosaik}~\cite{Raub2021}.}
\label{sametimeloops_example}
\end{figure}

Another important feature to allow hybrid co-simulation is the superdense time model, which allows multiple simulation steps to happen at the same simulation time as proposed in \cite{Cremona2017}.
This is shown in Figure~\ref{sametimeloops_example}, where the x-axis represents the simulation time and multiple data exchanges happen within the first step between the simulators A and B.
During this loop between simulators A and B, the simulation time is not increasing, and thus it is also called \textit{same time loop}.
With superdense time, phenomena at time scales much shorter than the scope of the simulation can be resolved instantaneously, e.g., the convergence of electronic controllers, negotiation between agents, or convergence between parts of large physical systems. This can be especially useful for identification of consistent initial states in a simulation.
\par
To implement the superdense time feature in \textsc{mosaik}, two simulation components have to be connected in a loop and one of these connections has to be defined as weak (\texttt{weak=True}).
To activate (and also stay in) a same time loop, a simulator has to provide the corresponding attributes via the \texttt{get\_data} function and setting the output time to the current step time. 
To end the same time loop, the output time has to be set to a time later than the current step time or no data has to be provided for attributes in the \texttt{get\_data} return dictionary.
For the iteration of same time loops between simulation components, a limit can be set in the instantiation of \texttt{mosaik.scenario.World} via the parameter \texttt{max\_loop\_iterations} to avoid infinite cycles, which is set to 100 by default.
A tutorial for same time loops in \textsc{mosaik} can be found in the documentation\footnote{\url{https://mosaik.readthedocs.io/en/latest/tutorials/sametimeloops.html}}.
\subsection{External events} 
\begin{figure}
\centering
\includegraphics[trim=8cm 4cm 8cm 4cm,clip, width=0.45\textwidth]{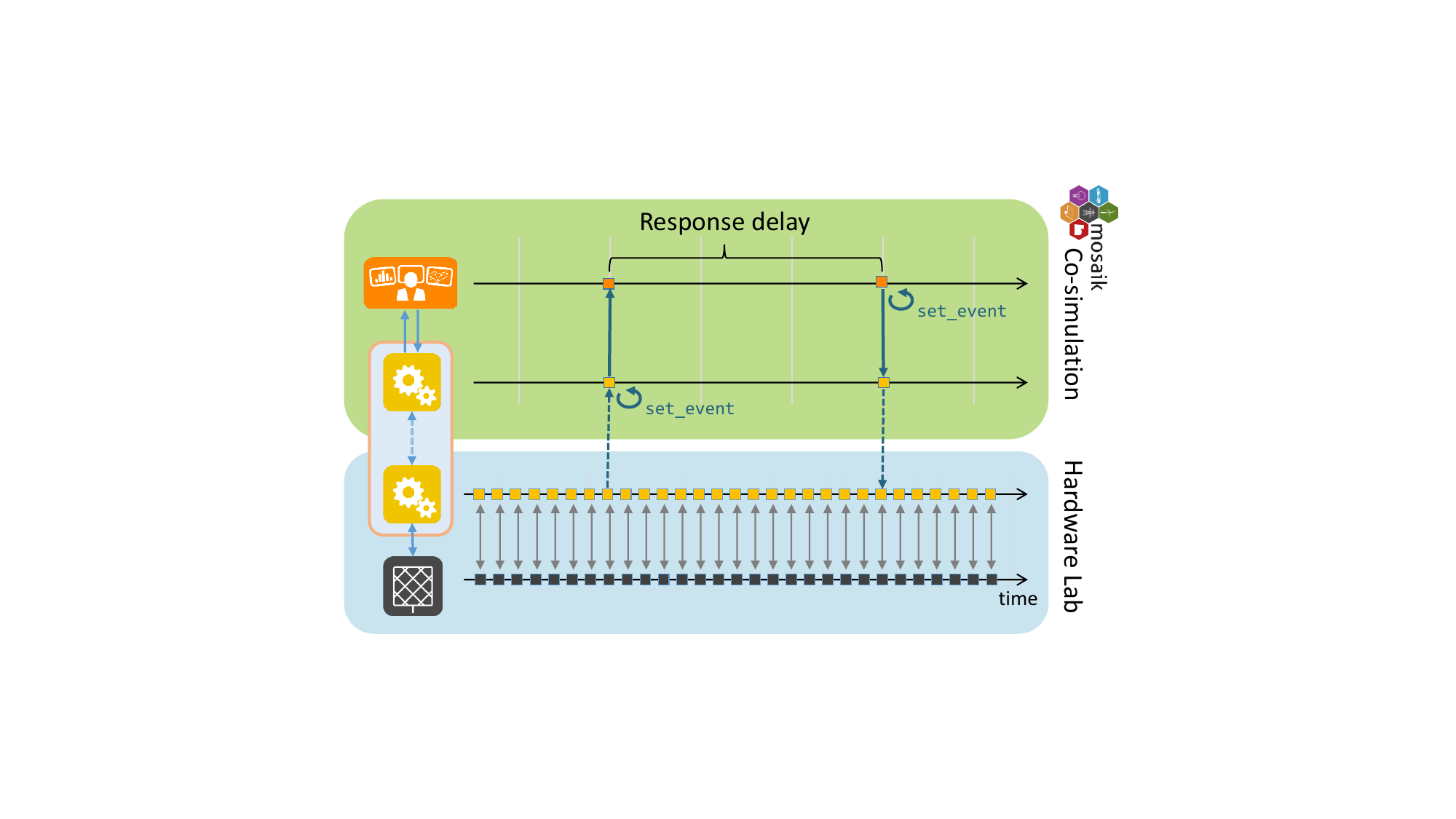}
\caption{Use Case: Human in the Loop simulation~\cite{Raub2021}.}
\label{set-event-use-case}
\end{figure}

The extended scheduling algorithm also supports external events (e.g., provided via a GUI), so that an unforeseen interaction of a model with an external system can be integrated in soft real-time simulations that are synchronized based on wall clock time \cite{Buscher2014,Rehtanz2016}. These external events can be implemented via the asynchronous \texttt{set\_event} function to trigger the stepping of a simulator within the subsequent time. For instance, this feature can be used for Human-in-the-Loop simulations to support human interactions, as shown in Figure \ref{set-event-use-case}. In this example, an operator interacts with a hardware laboratory simulation (e.g., SESA Lab\footnote{\url{https://www.offis.de/en/living-labs/sesa.html}} via a SCADA interface) to be able to control parts of the system. These control actions are passed as external events to a controller in \textsc{mosaik}, which forwards these events to its represented controller in the laboratory, where the control actions take place.
A tutorial for external events in \textsc{mosaik} can be found in the documentation\footnote{\url{https://mosaik.readthedocs.io/en/latest/tutorials/set-external-events.html}}.

\section{Future Work}\label{sec:future_work}
As described in the previous sections, the new features of \textsc{mosaik} 3.0 for hybrid co-simulation enable the implementation of new use cases, but also open up new opportunities for further extension and improvement of \textsc{mosaik}, which is planned in the future.
\par
\textsc{mosaik} already provides an adapter to integrate FMUs based on the FMI 2 specification into simulations\footnote{\url{https://gitlab.com/mosaik/components/mosaik-fmi}}. 
The new features of the FMI 3 release~\cite{FMI3} will be evaluated and aligned with the new \textsc{mosaik} discrete event capabilities to make use of these features.
\par
The new superdense time feature in \textsc{mosaik} allows to investigate new research questions, e.g., the effects of negotiation of agents or convergence of controllers in the initialization of co-simulation.
However, simulation with changing resolutions of simulators caused by the new event-based scheduling may also have an impact on the uncertainty of results.
Thus, approaches for integration of uncertainty quantification and propagation in \textsc{mosaik} will be extended to further investigate its effects~\cite{Steinbrink2017}.
\par
For standard compliant communication in real-time simulations, such as OPAL RT or RTDS, the IEC 60870-5-104 protocol (also known as 104 protocol or IEC 104) is commonly used. To implement the 104 protocol in \textsc{mosaik}, a wrapper will be written based on the component API. Therefore, the \textit{hybrid} type is useful for regularly receiving messages while the \texttt{set\_event} function introduces external events. 
\par
\textsc{mosaik} 3.0 provides sophisticated features that allow the implementation of extended scenarios to analyze the impact of delays in the exchange of messages, e.g., when Multi-Agent-Systems (MAS) are investigated. MAS allow the coordination and integration of information systems as well as to represent coordination (behaviors) and control mechanisms. A MAS framework called mango\footnote{\url{https://gitlab.com/mango-agents/mango}} is being developed that implements modularity and autonomy between agents. 
Advanced research on message exchange between agents (e.g., investigating the impact of delays) can be conducted with the new superdense time and \texttt{max\_advance} features of \textsc{mosaik} 3.0.

Furthermore, activities are planned with regards to combined power and communication system co-simulations to analyze the dynamic interdependent behavior of both domains. This coupling enables to study new co-simulation scenarios including communication disturbances and active control (e.g., resilient altering routing or enabling/disabling infrastructure devices) of the communication infrastructure in large scale power system setups, such as changing the communication topologies to simulate attacks, faults and failures on the communication infrastructure in \textsc{mosaik}.

\section{Conclusion} \label{sec:conclusion} 
\textsc{mosaik} 3.0 extends \textsc{mosaik} 2 capabilities for co-simulation with improved event capabilities (e.g., for multi-agent or communication system simulations) and combines time-stepped and discrete event simulations.
As part of the new features of \textsc{mosaik} 3.0, we have introduced changes to the component API, including the new simulator component type, the time parameter \texttt{max\_advance}, and the definition of a global time resolution. In addition, we have outlined some enhancements to the scheduling algorithm that allow to model superdense time and to set external events for Human-in-the-Loop simulations. We have shown that the approach of hybrid co-simulation, that embraces the combination of continuous time and discrete event models, can improve the simulation and analysis of complex CPES. Thus, the support of new standards such as FMI 3, the 104 protocol, and the new superdense time and external events features enable the investigation of new research questions in future scenarios.

\section*{Acknowledgment}
The concept of \textsc{mosaik} 3.0 was mainly developed by Thomas Raub and Reef Eilers, while the implementation was done by Thomas Raub.

\bibliographystyle{IEEEtran}
\bibliography{refs}

\end{document}